\documentclass[graybox]{svmult}%

\makeindex             

\begin{document}

\title*{A 2D field theory  equivalent to 3D gravity with no cosmological constant}
\author{Glenn Barnich, Andr\'es Gomberoff  and Hern\'an A. Gonz\'alez}
\institute{Glenn Barnich \at Physique Th\'eorique et Math\'ematique, Universit\'e Libre 
de Bruxelles and International Solvay Institutes, Campus Plaine C.P. 231, B-1050 Bruxelles, Belgium, \email{gbarnich@ulb.ac.be}
\and Andr\'es Gomberoff   \at Departamento de Ciencias F\'{\i}sicas, Universidad Andres Bello, Av. Rep\'{u}blica 252, Santiago, Chile, \email{agomberoff@unab.cl},  \and Hern\'an A. Gonz\'alez   \at Departamento de F\'{\i}sica, P. Universidad Cat\'olica de
Chile, Casilla 306, Santiago 22, Chile. \email{hdgonzal@uc.cl}} 
%
%
\maketitle

\abstract*{In (2+1) space-time dimensions the Einstein theory of gravity has no local degrees of freedom. In fact, in the presence of a negative cosmological term,  it is described by a (1+1) dimensional theory living on its boundary: Liouville theory.  It is invariant under the action of the two-dimensional conformal group, which, in the gravitational context, corresponds to the asymptotic symmetries of asymptotically AdS geometries.    
In the flat case,  when the cosmological term is turned off, a theory describing gravity at the boundary is absent. In this note we show that, in the Hamiltonian setup, such a theory may be constructed. The theory is BMS$_3$  invariant, as it should, corresponding to the asymptotic symmetry group of an asymptotically flat spacetime.}

\abstract{In (2+1) space-time dimensions the Einstein theory of gravity has no local degrees of freedom. In fact, in the presence of a negative cosmological term,  it is described by a (1+1) dimensional theory living on its boundary: Liouville theory.  It is invariant under the action of the two-dimensional conformal group, which, in the gravitational context, corresponds to the asymptotic symmetries of asymptotically AdS geometries.    
In the flat case,  when the cosmological term is turned off, a theory describing gravity at the boundary is absent. In this note we show that, in the Hamiltonian setup, such a theory may be constructed. The theory is BMS$_3$  invariant, as it should, corresponding to the asymptotic symmetry group of an asymptotically flat spacetime.}

\section{Introduction}
\label{sec:1}

In the last 30 years, gravitational theories in 2+1 spacetime dimensions have attracted much attention. A celebrated advance in this field is due to Brown and Henneaux in  \cite{Brown:1986nw}, where they found that asymptotically  AdS spacetimes have more symmetries than one expects. Instead of being symmetric under the $SO(2,2)$ group as one may have guessed, they turned out to be invariant under the whole conformal group in 2 dimensions, whose algebra gets centrally extended. This result is reminiscent from what was obtained in the study of asymptotically flat spacetimes where the group of asymptotic symmetries is much bigger than the Poincar\'e group one would expect. It is the infinite dimensional BMS group \cite{Sachs:1962zza,Bondi:1962px,Ashtekar:1996cd,Barnich:2006av}.   

Ten years after it was shown that, in  the presence of a negative cosmological term,  Einstein-Hilbert gravity is described by  Liouville theory \cite{Coussaert:1995zp}, which is known to be conformally invariant (see, for instance \cite{Seiberg:1990eb}). 

In the present talk, which is based on the work originally published in \cite{Barnich:2012rz}, we are going to show how we may take the limit of vanishing cosmological constant so that a field theory of asymptotically flat gravity is obtained. Although this procedure may appear to be trivial, it turns out that the limit is not well defined in the Lagrangian action when keeping a finite value of Newton's constant $G$. We will show, however, that in the Hamiltonian formulation a well defined limit may be taken for any value of $G$. 

\section{Liouville Theory and Gravity in 3D spacetime}

Liouville theory is defined by the action on the
Minkowskian cylinder with time coordinate time  $t$,
angular coordinate $\phi\in [0,2\pi)$ 
\begin{equation}
\label{la}
I[\varphi]=\int dt d\phi \, \Big(\frac{1}{2}\dot{\varphi}^2
-\frac{1}{2l^2}{\varphi'}^2+\frac{\mu}{2\gamma^2} 
e^{\gamma\varphi}\Big).
\end{equation}
The action has three independent parameters,  namely $\gamma$, $\mu$ and $l$.
However, $\mu$ is irrelevant in the sense that with a redefinition of the field $\varphi\rightarrow\varphi +$const., its value may be shifted to any non-zero value. 

The theory is equivalent to (2+1)-dimensional gravity \cite{Coussaert:1995zp}, when its constants are related to the gravitational ones by, 
\begin{equation}
\label{gv}
G=\frac{\gamma^2l^2}{32\pi}, \ \ \ \Lambda=-\frac{1}{l^2},
\end{equation}
where $\Lambda$ is the cosmological constant and $G$ is Newton's constant.

Liouville theory (\ref{la}) is known to be invariant under the conformal group. Properly normalized, the corresponding Virasoro algebra gets a central extension \cite{Seiberg:1990eb},
\begin{equation}\label{central}
c=\frac{48\pi}{\gamma^2 l}.
\end{equation}
Written in terms of the gravitational parameters (\ref{gv}) this is precisely the Brown-Henneaux central charge, $c_{BH}= 3l/2G$. 

It is clear, by inspection of (\ref{la}), that the limit of vanishing cosmological constant is not immediate if one wishes to keep $G$ finite.  Taking, $l\rightarrow\infty$ keeping $\gamma$ and $\mu$ finite one obtains
\begin{equation}
\label{lal}
I[\varphi]=\int dt d\phi \, \Big(\frac{1}{2}\dot{\varphi}^2
+\frac{\mu}{2\gamma^2} 
e^{\gamma\varphi}\Big).
\end{equation}
However this leads to $G \rightarrow\infty$ and a vanishing central charge $c=0$. 
The resulting theory  is invariant under bms$_3$ transformations, but it is not (2+1)-dimensional gravity. We will show in the next section how the limit may be taken using the Hamiltonian formulation keeping $G$ finite. We will also show  why this limit may not be taken in the Lagrangian formulation.

 \section{Hamiltonian formulation and the flat limit}

The Hamiltonian action of Liouville theory is,
\begin{equation}
\label{ha}
I[\varphi,\pi]=\int dt d\phi\, \Big(\pi \dot{\varphi}
-\frac{1}{2}\pi^2 - \frac{1}{2l^2}{\varphi'}^2-\frac{\mu}{2\gamma^2} 
e^{\gamma\varphi}\Big).\label{a}
\end{equation}
When minimizing the action with respect to the canonical momentum $\pi$ one obtains
\begin{equation}\label{mom}
\pi=\dot \varphi,
\end{equation}
as expected. One may recover the Lagrangian action (\ref{la}) by replacing the momentum   (\ref{mom})  
and putting it back in   (\ref{ha}).  

As is well-known, Liouville theory is invariant under two-dimensional
conformal transformations\cite{Seiberg:1990eb}.  In the Hamiltonian framework they are generated by
charges satisfying  a centrally extended conformal algebra, with a central charge given by  (\ref{central}). We are going to skip the derivation in this note. It is a well known result, and  a derivation using the notation and normalizations used here may be found in \cite{Barnich:2012rz}. 

We now study the $l\rightarrow\infty$ in the Hamiltonian version of the theory (\ref{ha}).
We may first consider the limit discussed at the end of Sec. 2, which leads us to a theory of vanishing central charge, of no use for gravity unless one wishes to study some strong coupling limit in which  $G\rightarrow\infty$ is relevant.  Doing so, the third term of (\ref{ha}) drops out. Varying the action respect to $\pi$ one again obtains (\ref{mom}), which once inserted back in the action, produces (\ref{lal}).  

In the Hamiltonian version, however, there is a second way of proceeding with the flat limit,  leading us to a theory which is equivalent to (2+1)-dimensional gravity for generic, finite $G$.
We first rescale the field and its
momentum through,
$
\varphi = l\Phi,$ $\pi=\Pi/l, 
$ and define \(\beta=\gamma l, \nu= \mu l^2\). We obtain,
\begin{equation}
  \label{Hgon05}
 I[\Phi,\Pi]=\int dt d\phi\, \Big( \Pi \dot{\Phi}-\frac{1}{2l^2}\Pi^2-
\frac{1}{2}\Phi'^2-\frac{\nu}{2\beta^2} e^{\beta\Phi}\Big). 
\end{equation}
Since the rescaling of variables is a canonical transformation, the 
Poisson algebra of the conformal group keeps its form and the central charge does not change.
We now take the limit $l\rightarrow\infty$ keeping $\beta$ and $\nu$ fixed. Note that, in this case, the 
second term in the action is similar to that of a particle of mass $l$, so that the limit mimics the one  of an ultra-massive particle,
\begin{equation}
  \label{final}
 I[\Phi,\Pi]=\int dt d\phi\, \Big( \Pi \dot{\Phi}-
\frac{1}{2}\Phi'^2-\frac{\nu}{2\beta^2} e^{\beta\Phi}\Big). 
\end{equation}
This action has no Lagrangian counterpart. Varying with respect to $\Pi$ gives no algebraic equation for it, and therefore the action cannot be reduced to a Lagrangian, second order form. The field $\Pi$ is now a Lagrangian multiplier. The constant $G$ is kept finite, because $\beta=\sqrt{32\pi G}$ is held fixed in the limit. The centrally extended Virasoro algebra becomes the centrally extended BMS$_3$ algebra in the way it was first found in \cite{Barnich:2006av}. Note that here there is a subtlety. The generators of BMS$_3$ must be properly rescaled before taking the limit, so that the central extension becomes proportional to $c/l$, which is finite in the limit as one may see from (\ref{central}) (see also\cite{Barnich:2012rz} for details). 
 
 The action (\ref{final}), with $\beta=\sqrt{32\pi G}$ and arbitrary $\nu$ is equivalent to Einstein gravity with no cosmological constant in the same way Liouville theory is when the cosmological constant is turned on\footnote{Note however that, as discussed in \cite{Henneaux:1999ib}, this equivalence generally holds only up to zero modes.}. The theory is invariant under the BMS$_3$ group, as it must be, because it is the asymptotic symmetry group of  asymptotically flat 3-dimensional gravity.  The particular form the fields are transformed by the group may be found in \cite{Barnich:2012rz}.
\begin{acknowledgement}
The work of G.B.~is supported in part by the Fund for Scientific
Research-FNRS (Belgium), by the Belgian Federal Science Policy Office
through the Interuniversity Attraction Pole P6/11, by IISN-Belgium, by
``Communaut\'e fran\c caise de Belgique - Actions de Recherche
Concert\'ees'' and by Fondecyt Projects No.~1085322 and
No.~1090753. The work of A.G.~was partially supported by Fondecyt
(Chile) Grant \#1090753. H.G.~thanks Conicyt for financial support.
\end{acknowledgement}

\end{document}